\documentclass[aps,prb,twocolumn,groupedaddress]{revtex4-2}

\usepackage[T1]{fontenc}

\usepackage{amsmath}
\usepackage{amssymb}
\usepackage{bbm}

\usepackage{xspace}
\usepackage{color}
\usepackage{graphicx}
\usepackage{ulem} 
\usepackage{placeins}

\usepackage[colorlinks=true,linkcolor=blue,citecolor=blue,urlcolor=blue]{hyperref}

\graphicspath{{.}{Bilder/}}

\newcommand*{\imag}{\mathrm{i}}
\newcommand*{\HgTe}{\mbox{HgTe}\xspace}
\newcommand*{\BiSe}{\mbox{Bi\textsubscript{2}Se\textsubscript{3}}\xspace}

\newcommand{\ie}{\textit{i.e.}\xspace} 	
\newcommand{\eg}{\textit{e.g.}\xspace} 	

\begin{document}


\title{Crossed Andreev reflection in topological insulator nanowire T-junctions}


\author{Jacob Fuchs$^1$}
\author{Michael Barth$^1$}
\author{Cosimo Gorini$^1$}
\author{\.{I}nan\c{c}  Adagideli$^{2,3}$}
\author{Klaus Richter$^1$}%
\affiliation{%
    $^1$Institut  f\"ur  Theoretische  Physik,  Universit\"at  Regensburg,  93040  Regensburg,  Germany \\
    $^2$Faculty of Engineering and Natural Sciences, Sabanc\i{} University, 34956 Orhanl\i{}-Tuzla, Turkey \\
    $^3$Faculty of Science and Technology and MESA+ Institute for Nanotechnology,
University of Twente, 7500 AE Enschede, The Netherlands.}
%


\date{\today}

\begin{abstract}
We numerically study crossed Andreev reflection (CAR) in a topological insulator nanowire T-junction
where one lead is proximitized by a superconductor.
We perform realistic simulations
based on the 3D BHZ model
and compare the results
with those from an effective 2D surface model,
whose computational cost is much lower.
Both approaches show that CAR should be clearly observable
in a wide parameter range,
including perfect CAR in a somewhat more restricted range.
Furthermore, it can be controlled by a magnetic field
and is robust to disorder.
Our effective 2D implementation allows to model systems of micronsize,
typical of experimental setups,
but computationally too heavy for 3D models.
\end{abstract}


\maketitle

\section{Introduction\label{sec:introduction}}

The combination of superconductors (S) with materials in the normal (N) conducting state led to the discovery of many interesting physical effects \cite{Tinkham-superconductivity-1996},
a notable one being Andreev reflection (AR) \cite{1964-JETP-Andreev}.
In this process an incoming electron from the N-contact is reflected as a hole by forming a Cooper pair in the superconductor.
In the presence of a second N-contact,
the outgoing hole can either leave through the same normal lead as the incoming electron,
or through the other and spatially separated one.
The second process is called crossed Andreev reflection (CAR)
and amounts to the formation of a Cooper pair from two electrons from different leads.
CAR is particularly interesting because of its non-local character.
In fact, non-locality can be exploited to generate entanglement via CAR's reciprocal process, \ie the ``splitting'' of a Cooper pair into two entangled electrons
leaving the system through different contacts.  A Cooper pair splitter is a three-terminal setup
with one superconducting and two normal contacts, has been investigated both theoretically (e.g. Refs.~\cite{2000-ApplPhysLett-Deutscher_Feinberg,2001-EurPhysJB-Lesovik_Martin_Blatter,2001-PhysRevB-Recher_Sukhorukov_Loss,2010-PhysRevB-Haugen_et_al})
and experimentally (e.g. Refs.~\cite{2009-Nature-Hofstetter_et_al,2010-PhysRevLett-Herrmann_et_al}).
However, CAR generally competes with normal electronic transmission (T), where an electron 
is directly transferred from one normal lead to the other, bypassing the superconducting contact.
Indeed, T usually dominates CAR.
The dominating process can be identified with the nonlocal conductance,
to which T and CAR contribute with opposite signs.


Topological insulators (TIs) exhibit a number of peculiar transport properties,
\eg surface or edge transport robust to disorder \cite{2010-PhysRevLett-Bardarson_Brouwer_Moore}, helical edge modes in 2D TIs, \cite{2013-PhysStatusSolidiB-Tkachov_Hankiewicz}
topological superconductivity and Majorana modes when proximitized by a superconductor \cite{2010-RevModPhys-Hasan_Kane}.
Despite beeing forbidden in 2D TIs due to the helical nature of the (edge) transport channels \cite{2010-PhysRevB-Adroguer_et_al},
signs of CAR were reported in the presence of magnetic ordering \cite{2008-PhysRevLett-Nilsson_Akhmerov_Beenakker},
couplings between the edges \cite{2011-PhysRevB-Chen_et_al,2013-PhysRevLett-Reinthaler_Recher_Hankiewicz},
odd-frequency triplet superconductivity \cite{2015-PhysRevB-Crepin_Burset_Trauzettel,2018-PhysRevB-Fleckenstein_Ziani_Trauzettel},
or when the system is arranged in a bipolar setup \cite{2008-PhysRevLett-Cayssol,2018-PhysRevLett-Breunig_Burset_Trauzettel,2017-PhysRevB-Islam_Dutta_Saha}.

\begin{figure}
  \centering
  \includegraphics[width=0.6\linewidth]{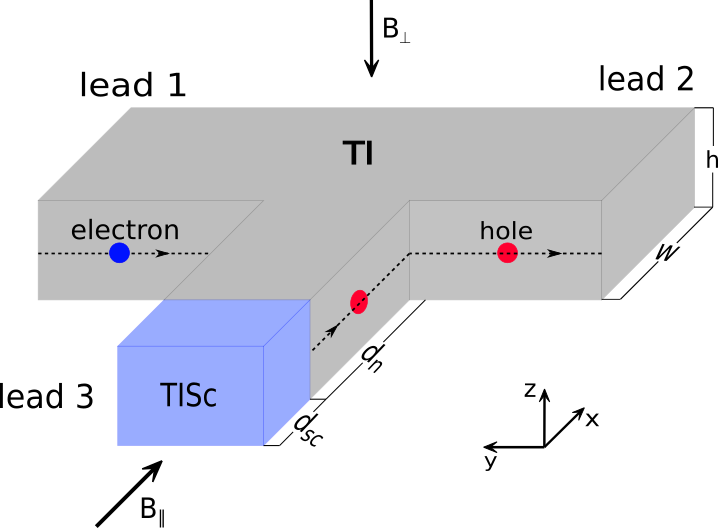}
  \caption{Schematic illustration of the suggested setup. The system consists of a junction of three TI nanowires where one of them is proximitized by a normal s-wave superconductor. Furthermore two magnetic field components are shown, a perpendicular one to induce chiral edge states and an axial magnetic field to tune the local AR at the interface.}
  \label{fig:t-junction}
\end{figure}

In this paper we propose a Cooper pair splitter setup based on a 3D TI T-junction,
whose third lead is a normal-superconductor (NS) junction, see Fig.~\ref{fig:t-junction}.
In a simpler 2-terminal geometry, \ie~a straight 3D TI nanowire, 
the NS contact allows switching between (local) Andreev reflection (AR) and electron reflection (R)
by tuning a coaxial magnetic field \cite{2014-PhysRevLett-deJuan_Ilan_Bardarson}.
Once embedded in our setup, the same NS contact allows nonlocal switching between T and CAR by tuning $\vec{B}_\parallel$,
which plays the role of the coaxial field of the 2-terminal configuration.
Furthermore, we will see that CAR is expected for a wide parameter range,
suggesting the feasibility of our device with current technology.

The paper is straightforwardly organized:
In Sec.~\ref{sec:t-junction} we describe the T-junction device and its working principles;
Sec.~\ref{sec:methods} introduces the model and the numerical methods we base our simulations on, while
Sec.~\ref{sec:results} discusses the numerical results.  We conclude with a brief summary in Sec.~\ref{sec:conclusion}.
Certain technicalities can be found in the Appendices.
\FloatBarrier

\section{T-junction device\label{sec:t-junction}}

\subsection{Previous work: NS junction\label{sec:ns-junction}}

It is useful to start by recalling the main magnetotransport characteristics of the basic NS junction.

In 3D TI nanowires,
a intrinsic Berry phase of $\pi$ leads to a gapped spectrum
\cite{2010-PhysRevLett-Zhang_Vishwanath,2010-PhysRevLett-Bardarson_Brouwer_Moore,2010-PhysRevLett-Ostrovsky}.
When a magnetic field $B_{\parallel}$ coaxial with the wire
is switched on, the Berry phase competes with the resulting Aharonov-Bohm one, \ie with the flux $\phi$ through the nanowire cross-section.  The latter tunes the spectrum and makes it gapless for $\phi=\phi_0/2$, with $\phi_0 = h/e$ the flux quantum.
At this half-integer value of $\phi$, topological superconductivity is induced in the system by proximity with a trivial superconductor
\cite{2011-PhysRevB-Cook_Franz,2012-PhysRevB-Cook_Vazifeh_Franz}, reverting back to triviality for other values of the flux.
Topological and trivial regimes can be distinguished by the 2-terminal conductance of a single NS junction \cite{2014-PhysRevLett-deJuan_Ilan_Bardarson}.
The two regimes remain discernible when a strong perpendicular mangetic field $B_{\perp}$ is applied to the normal half of the system, provided it is screened on the superconducting side.
Then, Landau levels and chiral edge states emerge
\cite{%
    2009-PhysRevLett-Lee,%
    2011-PhysRevB-Vafek,%
    2012-JPhysCondensMatter-Zhang_Wang_Xie,%
    2014-PhysRevB-Brey_Fertig%
}.
In the single-mode regime,
there is only one incoming channel on the one edge
and one outgoing channel on the other edge.
Incoming electrons are
either perfectly reflected in the trivial regime
(leading to a NS conductance of $0$)
or perfectly Andreev reflected in the topological regime
(such that the NS conductance is $2 e^2 / h$).
This effect arises from the flux $\phi=\phi_0/2$ threaded through the superconducting lead, introducing a vortex at the NS-interface \cite{2012-PhysRevB-Cook_Vazifeh_Franz}. 
The vortex modifies Andreev reflection \cite{2014-PhysRevLett-deJuan_Ilan_Bardarson},
by determining the matching angular momenta of electron and hole modes in the single-mode regime. 
In this regime a robust Andreev reflection signature can be obtained, which in turn can be used for a robust CAR process.

\subsection{Mechanics of the T-junction\label{sec:mechanics}}

Incoming and outgoing quasiparticles are spatially separated,
travelling along chiral edge states on opposite sides of the wire.  To split them into separate leads,
another nanowire is attached to get the T-shaped device
shown in Fig.~\ref{fig:t-junction}.
In our case wires have a rectangular cross section,
so that, strictily speaking, the edge states are actually side surface states.
The magnetic field $B_{\perp}$ is perpendicular to the full
T-junction structure, which is then as a whole in the quantum Hall regime, with Landau levels on the top and bottom T-shaped surfaces.  The incoming and outgoing channels at the NS interface thus spatially separate, running into/coming from different normal leads.  This way AR turns into CAR. 
The parallel magnetic field $B_{\parallel}$
now controls not only the AR at the NS interface,
but also the CAR of the entire device:
For $\phi = 0$,
we expect normal reflection at the NS interface
and electron transmission (T) in the entire device,
while for $\phi = \phi_0/2$
Andreev reflection
and crossed Andreev reflection, respectively.

Note that $B_{\parallel}$ should not be strong enough to push the side T-junction surfaces into the quantum Hall regime.
This condition is easily met in practice, as the height of typical nanowires is considerably smaller than their width.
(c.f. Ref.~\cite{2018-PhysRevB-Ziegler_et_al}, for example).

\section{Methods\label{sec:methods}}

\subsection{Surface model\label{sec:surface-model}}

Due to the bulk being insulating,
the system can be modeled by describing the surface states with a 2D Dirac Hamiltonian \cite{%
    2010-PhysRevLett-Bardarson_Brouwer_Moore,%
    2010-PhysRevLett-Zhang_Vishwanath,%
    2014-PhysRevB-Brey_Fertig,%
    2018-PhysRevB-Ziegler_et_al%
}.
We use an approach similar to Ref.~\cite{2014-PhysRevB-Brey_Fertig}
which can be applied to arbitrarily shaped devices.
The Hamiltonian of a surface with normal vector $\vec{n}$ reads
\begin{equation}
    \label{eq:H-surface}
    H_{\vec{n}} = \hbar v_F ( \vec{\sigma} \times \vec{k} ) \cdot \vec{n} - \mu
\end{equation}
with $\vec{\sigma}$ the vector of Pauli matrices
and $\mu$ the chemical potential.
The wave functions of the different surfaces
have to be matched at the edges, $\psi_1 = U \psi_2$,
where $U$ is the appropriate spin rotation
\footnote{%
    This is the special case of \cite{2014-PhysRevB-Brey_Fertig}
    for $A_1 = A_2 = \hbar v_F$ and $D_1 = D_2 = 0$.%
}.
For the Fermi velocity,
we use $\hbar v_F = 0.41\,\mathrm{eV\,nm}$
for \BiSe \cite{2009-NatPhys-Zhang_et_al},
and $\hbar v_F = 0.33\,\mathrm{eV\,nm}$
for \HgTe \cite{2018-PhysRevB-Ziegler_et_al}.
Superconductivity is modelled using the Bogoliubov-de Gennes formalism
(see sec.~\ref{sec:superconductivity}).
The magnetic field is described by the vector potential
\begin{align}
    \label{vector_potential}
    \vec{A}
    = \vec{A}_{\parallel} + \vec{A}_{\perp}
    = \begin{pmatrix}
        0 \\ - B_{\parallel} z / 2 \\ B_{\parallel} y / 2
    \end{pmatrix}
    + \begin{pmatrix}
    0 \\ - B_{\perp} x \\ 0
    \end{pmatrix}
    .
\end{align}
The origin of the coordinate system
is located at the NS interface
in the center of the nanowire.

For the numerical simulations,
we discretize the Hamiltonian to get a tight-binding model.
To deal with fermion doubling \cite{%
    1977-PhysRevD-Susskind,%
    1981-PhysLettB-Nielsen_Ninomiya,%
    1982-PhysRevD-Stacey%
},
we add the quadratic term
$(\vec{k}^2 - (\vec{k}\cdot\vec{n})^2) \, \vec{\sigma}\cdot\vec{n}$
to the Hamiltonian $H_{\vec{n}}$ of Eq.~\ref{eq:H-surface}.
The matching conditions $U$ between the surfaces
also enter the edge hoppings
as can be seen in App.~\ref{sec:matching}.
The magnetic field is introduced via Peierls substitution \cite{Peierls1933}
(see App.~\ref{sec:appendix_peierls}).

From the tight-binding model,
we calculate the scattering matrix
and transmission coefficients at zero energy
using \textsc{Kwant} \cite{%
    2014-NewJPhys-Groth_Wimmer_Akhmerov_Waintal,%
    2001-SIAMJMatrixAnalAppl-Amestroy_Duff_Excellent_Koster,%
    2019-ACMTransMathSoftw-Amestroy_Buttari_LExcellent_Mary%
}.
The conductances are given by \cite{1998-JPhysCondensMatter-Lambert_Raimondi}
\begin{align}
    \label{eq:G}
    G_{ba} &= \frac{\partial I_b}{\partial V_a} (V_a = 0), \\
    \label{eq:G_aa}
    G_{aa} &= \frac{e^2}{h} \left( N_a + T_{aa}^{\text{AR}} - T_{aa}^{\text{R}} \right), \\
    \label{eq:G_ba}
    G_{ba} &= \frac{e^2}{h} \left( T_{ba}^{\text{CAR}} - T_{ba}^{\text{T}} \right)
    , \quad \text{for $a \neq b$,}
\end{align}
where $I_b$ is the current from lead $b$ into the scattering region,
$N_a$ is the number of modes in lead $a$,
and $T_{aa}^{\text{R}}$, $T_{aa}^{\text{AR}}$, $T_{ba}^{\text{T}}$, and $T_{ba}^{\text{CAR}}$
are the transmission coefficients of
the electron reflection R (an electron from lead $a$ is reflected as electron),
the Andreev reflection AR (an electron from lead $a$ is reflected as hole),
the electron transmission T (an electron from lead $a$ is transmitted to lead $b$ as electron),
and the crossed Andreev reflection CAR (an electron from lead $a$ is transmitted to lead $b$ as hole).

\subsection{3D BHZ Model\label{sec:bhz-model}}

For our full 3D simulations we use the 3D BHZ Hamiltonian \cite{2009-NatPhys-Zhang_et_al,2010-Liu-PhysRevB.82.045122} in the basis
$\{ {|p1^+_z\uparrow\rangle}, {|p2^-_z\uparrow\rangle}, {|p1^+_z\downarrow\rangle}, {|p2^-_z\downarrow\rangle} \}$,
that is
\begin{align}\label{eq:H_BHZ}
H^{3D}=&\left(\epsilon(\vec{k})-\mu\right)\mathbbm 1_{4\times 4} \nonumber \\
&+
\begin{pmatrix}
M(\vec{k})    &  A_1k_z &        0         &           A_2k_- \\
A_1k_z &  -M(\vec{k})   &     A_2k_-        &     0           \\
0             &       A_2k_+    &    M(\vec{k})    &  -A_1k_z \\
A_2k_+         &        0       &  -A_1k_z  &   -M(\vec{k})
\end{pmatrix}.
\end{align}
Here
\begin{align}
\epsilon(\vec{k}) &= C + D_1k_z^2 + D_2(k_x^2+k_y^2), \\
M(\vec{k}) &= M - B_1k_z^2 - B_2(k_x^2+k_y^2), \\
k_\pm &= k_x \pm \imag k_y
.
\end{align}
We use \BiSe parameters, see Table \ref{table_bi2se3_params}.
\begin{table}
\begin{tabular}{ |l|l|l|l| }
\hline
$M= 0.28\,\mathrm{eV}$ & $A_1=2.2\,\mathrm{eV\,\mathring{A}}$ \\
\hline
$C_{\phantom{1}}=-0.0068\,\mathrm{eV}$ & $A_2=4.1\,\mathrm{eV\,\mathring{A}}$ \\
\hline
$B_1=10\,\mathrm{eV\,\mathring{A}^2}$ & $D_1=1.3\,\mathrm{eV\,\mathring{A}^2}$ \\
\hline
$B_2=56.6\,\mathrm{eV\,\mathring{A}^2}$ & $D_2=19.6\,\mathrm{eV\,\mathring{A}^2}$ \\
\hline
\end{tabular}
\caption{Hamiltonian parameters for \BiSe.}
\label{table_bi2se3_params}
\end{table}
Analogously to the 2D case the Hamiltonian is discretized, turning it into tight-binding form for implementation in \textsc{Kwant}. The cubic lattice has grid spacing $a=1\,\mathrm{nm}$, and the magnetic fields enters as before via Peierls substitution \cite{Peierls1933}. 
In the 3D model it is important for the axial magnetic field $B_\parallel$ to take into account the finite extension of the surface states into the bulk.
This is of relevance, as we want to have a flux of $\phi=\phi_0/2$ penetrating the NS-interface.
The penetration depth of the surface states defines an effective cross section area and therefore it is necessary to rescale the field strength \cite{2018-PhysRevB-Moors_et_al}.
In App.~\ref{sec:appendix_peierls} the modified hoppings are given for the vector potential defined in Eq.~\eqref{vector_potential}.
The orbital effect of the axial magnetic field is also considered inside the superconducting lead. 
The actual form of the vector potential becomes important in the case of superconducting systems, as it enters the gauge invariant phase difference and determines the supercurrent density in the system \cite{2019-PhysRevB.99.245408-Wimmer}.
In this work we adopt the choice of \cite{2019-SciPostPhys-deJuan_Bardarson_Ilan} and fixed the gauge to Eq.~\eqref{vector_potential}.

\subsection{Superconductivity\label{sec:superconductivity}}

Superconductivity is modelled by the Bogoliubov-de Gennes (BdG) Hamiltonian $H=\frac{1}{2}\Psi^\dagger\hat{H}\Psi$ with 
\begin{equation}
\hat{H}=
\begin{pmatrix}
H && \Delta(\vec{r})   \\
\Delta^*(\vec{r}) && -\mathbb{T}^{-1}H\mathbb{T}
\end{pmatrix}
\end{equation}
where $H$ is either the effective 2D Hamiltonian from Eq.~\eqref{eq:H-surface} or the full 3D Hamiltonian from Eq.~\eqref{eq:H_BHZ}.
The parameter $\Delta$ is the superconducting pairing potential, which depends on the spatial coordinate $\vec{r}$ and a complex phase $\chi$. The pairing potential is defined as
\begin{equation}
    \Delta(\vec{r}) = \Delta(r)\exp(\imag\chi(y,z))
\end{equation}
where
\begin{equation}
    \Delta(r) =
    \begin{cases}
        \Delta_0, & \text{$x < d_{sc}$} \\
        0, & \text{$x > d_{sc}$}
    \end{cases}
\end{equation}
with $\Delta_0 = 0.25\,\mathrm{meV}$ \cite{2014-PhysRevLett-deJuan_Ilan_Bardarson}.

Due to the fact that we are also considering the orbital effect of the axial magnetic field in the superconducting contact, the phase of the pairing potential will depend on the applied field strength and the spatial coordinates. 
Recall that a flux $\phi = \phi_0 / 2$ penetrating the superconducting wire cross-section is expected to introduce a vortex at the NS-interface. 
The vortex in our system is defined as
\begin{equation*}
    \chi(y,z) = \left \lfloor 2 \frac{\phi}{\phi_0} \right \rfloor
    \begin{cases}
        \arctan\left(\frac{z}{y}\right), &
        \arctan\left(\frac{z}{y}\right) > 0 \\
        2\pi+\arctan\left(\frac{z}{y}\right), &
        \arctan\left(\frac{z}{y}\right) < 0
    \end{cases}
\end{equation*}
with $\left \lfloor x \right \rfloor$ being the floor function.
For simplicity, and as we only consider surface states, we neglect the decrease of the pairing potential amplitude into the bulk due to the vortex.

\section{Results\label{sec:results}}

\subsection{Occurrence of CAR} \label{sec:occurence_of_CAR}

\begin{figure}
\includegraphics[width=\linewidth]{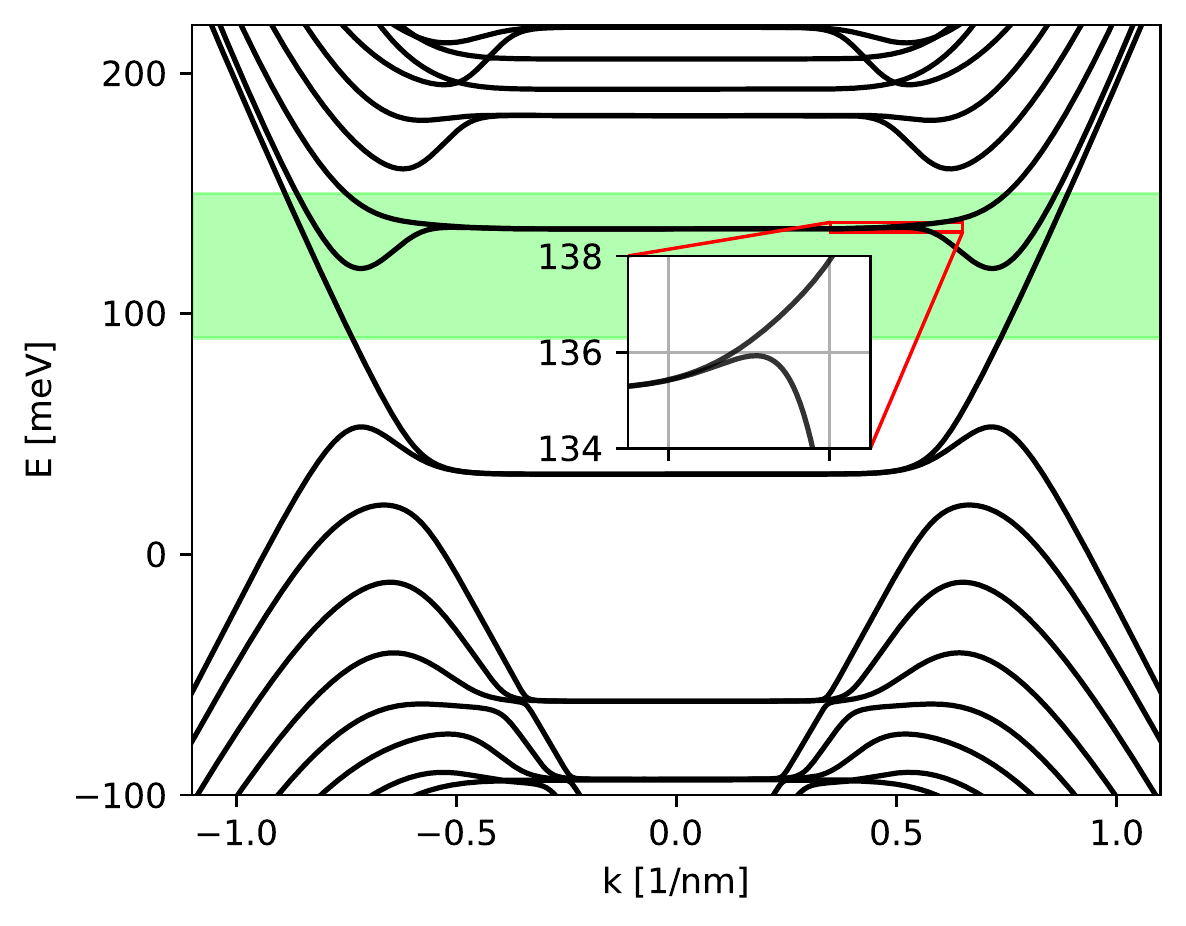}%
\caption{Bandstructure of a TI nanowire of width $w=50$ $nm$ and height $h=10$ $nm$ in a perpendicular magnetic field $B_\perp=20$ $T$ calculated for the 3D BHZ model. The asymmetry with respect to the zeroth Landau level stems from the anisotropy in the \BiSe crystal structure. The green shaded region corresponds to the energy range for our numerical transport calculations. This energy range corresponds to a peak in the transmission coefficients shown in Fig.~\ref{fig:bise-transmission}. \\
Inset: Zoom into the edge of the first Landau level.}
\label{fig:bands_B20}
\end{figure}

First, we simulate a \BiSe T-junction
with wires of width $w = 50\,\mathrm{nm}$, height $h = 10\,\mathrm{nm}$;
$d_{sc} = 1\,\mathrm{nm}$ and $d_n = 20\,\mathrm{nm}$ (see Fig.~\ref{fig:t-junction}).
These or larger dimensions are experimentally realizable \cite{2020-AdvancedElectronics-Rosenbach}.
We use $B_{\perp} = 20\,\mathrm{T}$
such that the wires are in the quantum Hall regime
(see Fig.~\ref{fig:bands_B20}).
Wires with larger diameters
allow for smaller magnetic fields $B_{\perp}$ (see below)
but are computationally more demanding.
Nevertheless, the transport signatures should not change qualitatively for larger systems. 

The sign of $B_{\perp}$ is chosen such that modes incoming from the left lead hit the NS interface,
whereas those from the right lead stay on the back side of the device,
and thus never reach the NS junction.
Since we focus on CAR, we restrict to incoming electrons from the left lead.

\begin{figure*}
\includegraphics[%
    width=\linewidth,%
]{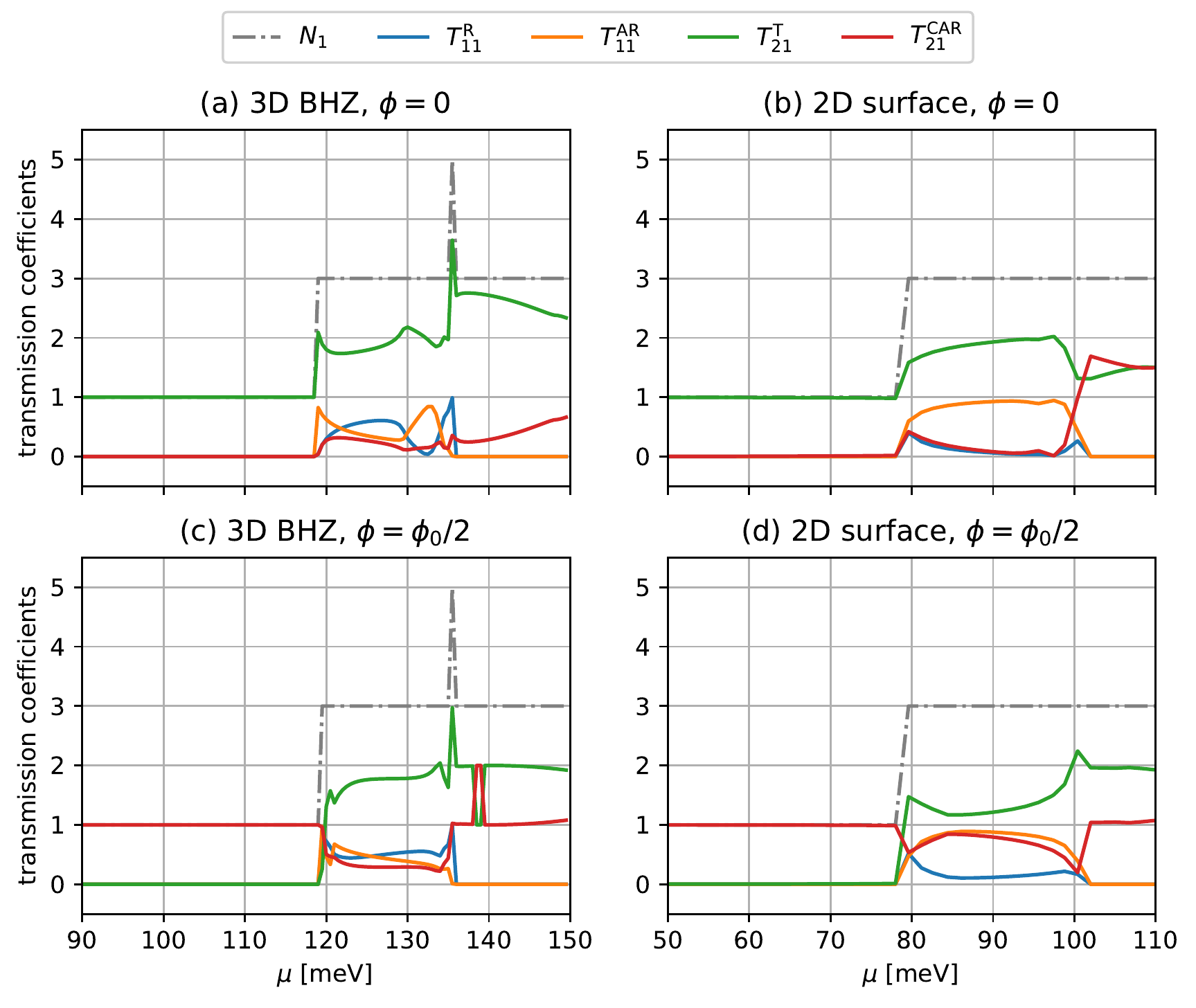}%
\caption{%
    Transmission coefficients
    of the T-junction device from fig.~\ref{fig:t-junction}.
    The height and width of the nanowires are
    $h = 10\,\mathrm{nm}$ and $w = 50\,\mathrm{nm}$,
    and the magnetic field is $B_{\perp} = 20\,\mathrm{T}$.
    \label{fig:bise-transmission}%
}
\end{figure*}

Without axial magnetic field ($B_{\parallel} = 0$),
there is perfect electron transmission and no CAR
for $\mu < 118\,\mathrm{meV}$
as can be seen in Fig.~\ref{fig:bise-transmission}(a).
At $\mu \approx 118\,\mathrm{meV}$,
an additional counter-propagating mode appears
\cite{2012-JPhysCondensMatter-Zhang_Wang_Xie},
coming from the small side minima of the Landau levels seen in Fig.~\ref{fig:bands_B20}.
Therefore, reflection processes (R and AR) become possible.
At $\mu \approx 135\,\mathrm{meV}$, the first Landau level is crossed.
The transmission peak and higher number of overall modes at $\mu \approx 135\,\mathrm{meV}$ are due to the Landau levels not being perfectly flat.
The 3D model anisotropy causes a small distortion of the Landau level dispersion in the $k$-region just before the strong side upward bending: the dispersion is slightly s-shaped -- see inset of Fig.~\ref{fig:bands_B20} --  resulting in a small energy range with 5 modes rather than 3.
This signature is absent in the 2D model, as the latter is isotropic.
For larger values of $\mu$, only T and CAR are possible.

Next, we use an axial magnetic field of $B_{\parallel} = 4.6\,\mathrm{T}$
to induce a flux $\phi \approx \phi_0 / 2$ through the NS interface
inducing a vortex (see Sec.~\ref{sec:superconductivity}).
The transmission coefficients are shown in Fig.~\ref{fig:bise-transmission}(c).
The single mode regime now exhibits perfect CAR in the energy range $\mu < 118\,\mathrm{meV}$.
Reflection processes appear as before only for $118\,\mathrm{meV} < \mu < 135\,\mathrm{meV}$
due to counter-propagating modes. 
CAR persists at higher energies, but T becomes dominant in that range.
The switching of T and CAR around $\mu \approx 138\,\mathrm{meV}$ is a numerical issue.  It appears because of an artificial mode mismatch at the NS interface between the superconducting lead, which hosts a vortex when $\phi=\phi_0 / 2$, and the nanowire surface states.  The flux enclosed by the latter is not exactly $\phi_0 / 2$, as the states extend a few sites into the 3D bulk.  The closer the value approaches the nominal $\phi_0/2$ in the superconducting lead, the smaller the numerical glitch.  

\begin{figure}
\includegraphics[%
    width=\linewidth,%
]{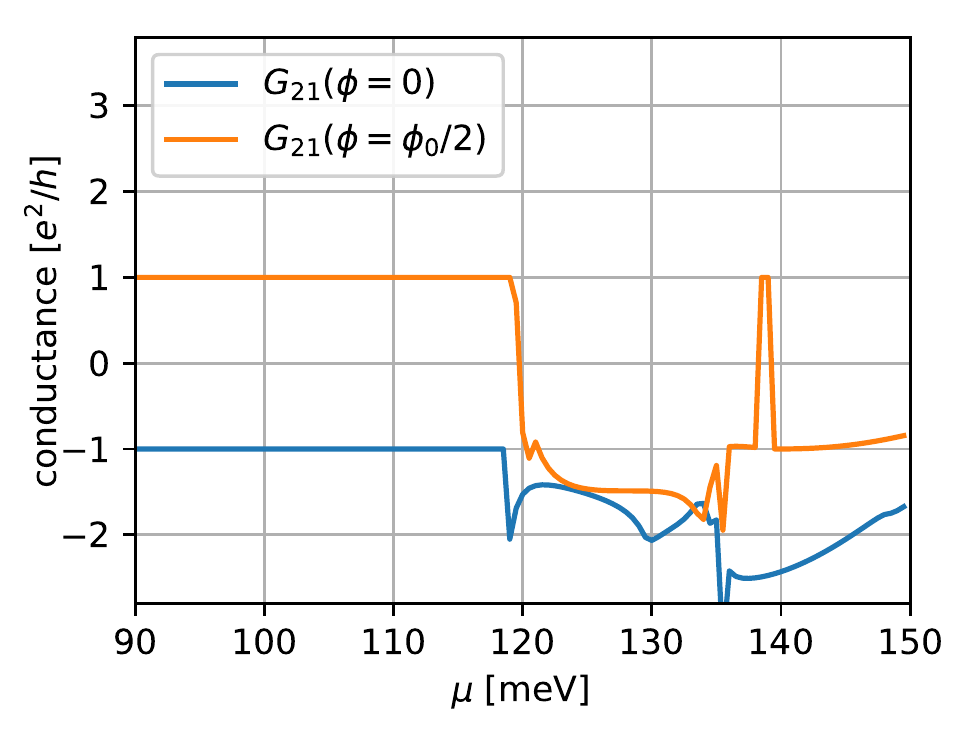}%
\caption{%
    Non-local conductance
    of the T-junction device from fig.~\ref{fig:t-junction}
    obtained from the 3D BHZ-model (see sec.~\ref{eq:H_BHZ}) for \BiSe
    (cf. fig.~\ref{fig:bise-transmission}(a,c)).
    \label{fig:bise-conductance}%
}
\end{figure}
Figure~\ref{fig:bise-conductance} shows the non-local conductances for the same system.
For $\phi \approx \phi_0 / 2$, the conductance is quantized and positive,
meaning that a voltage bias at the left lead
drives a current into the device and out to the right lead.
The edge states for higher energies
allow the conductance to become negative again.
For the applied perpendicular magnetic field strength it is more likely for those states to experience normal reflection at the NS interface.
Therefor the relative importance of CAR in the conductance drops.

The surface model results, Fig.~\ref{fig:bise-transmission}(b,d),
are in good agreement with those from the 3D BHZ model.
Note that there is a shift in the energy domain
since the surface model does not account
for the offset of the BHZ Hamiltonian.

\begin{figure}
\includegraphics[%
    width=\linewidth,%
]{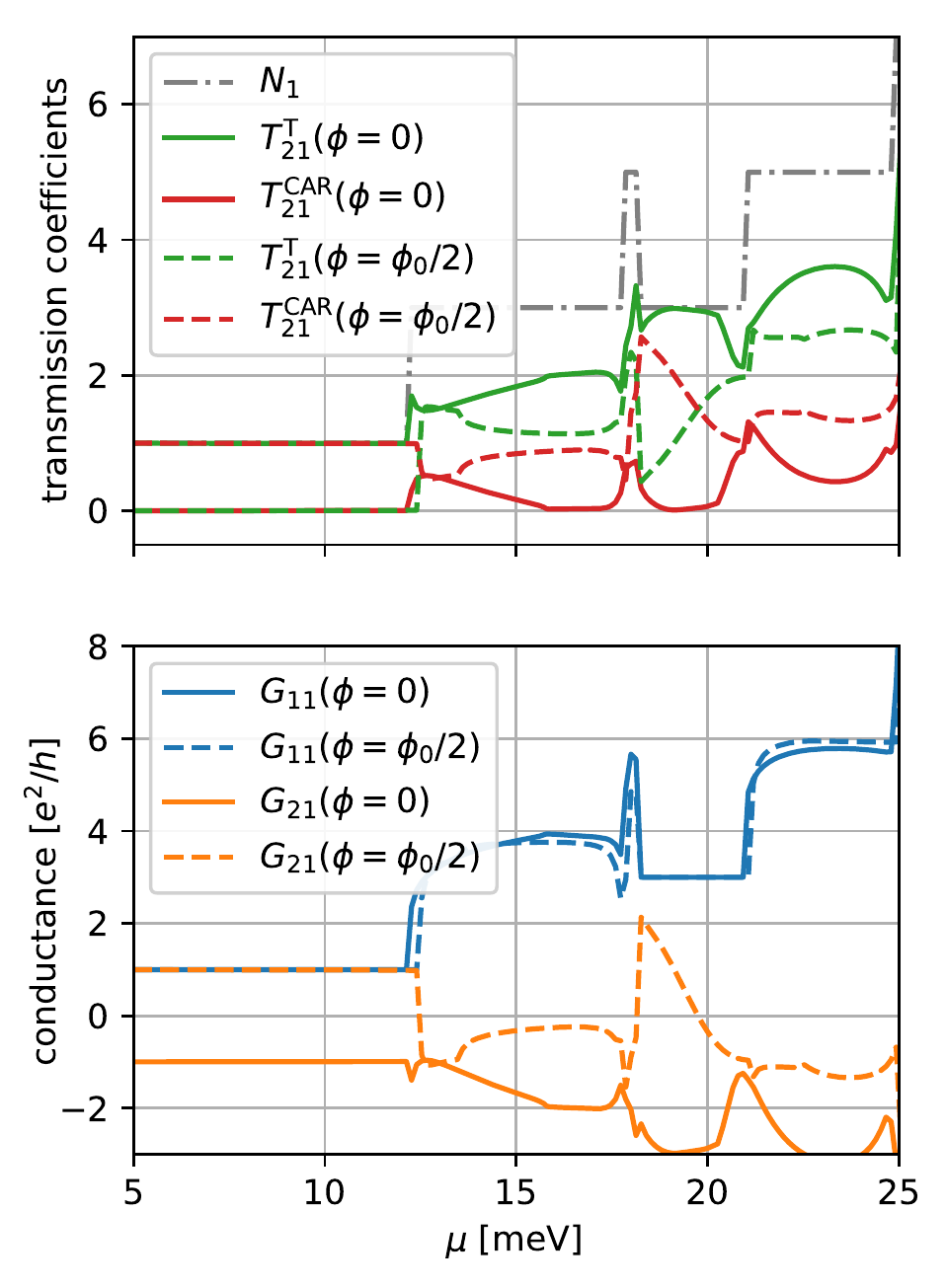}%
\caption{%
    Transmission coefficients and conductances
    of the T-junction device (see fig.~\ref{fig:t-junction})
    calculated with the surface model from Sec.~\ref{sec:surface-model}.
    The parameters are chosen to match the experimental values
    for \HgTe-nanowires from Ref.~\cite{2018-PhysRevB-Ziegler_et_al}:
    $h = 70\,\mathrm{nm}$, $w = 160\,\mathrm{nm}$,
    and $B_{\perp} = 1\,\mathrm{T}$;
    a vortex in the superconducting contact is present for $\phi = \phi_0/2$.
    \label{fig:surface-experiment}%
}
\end{figure}

Similar results are obtained for \HgTe nanowires,
see Fig.~\ref{fig:surface-experiment}.
The simulations are performed for wires of width $w = 160\,\mathrm{nm}$,
and height $h = 70\,\mathrm{nm}$, corresponding to the recent experimental sample sizes
\cite{2018-PhysRevB-Ziegler_et_al}.
In this system a perpendicular field $B_{\perp} = 1\,\mathrm{T}$
is enough to drive the system into the quantum Hall regime,
since the corresponding magnetic length
$l_B = \sqrt{\hbar / e B_{\perp}} \approx 26\,\mathrm{nm} \ll w$.


\subsection{Weak perpendicular magnetic field} 

Experimentally it can be quite challenging to tune the system close to the Dirac point and to access the single mode regime. 
For a clear and robust CAR signature it is desirable to operate the device in this energy range.  The requirement is however not necessary, as CAR signatures can be obtained also in other parameter regimes.

\begin{figure}
\includegraphics[width=\linewidth]{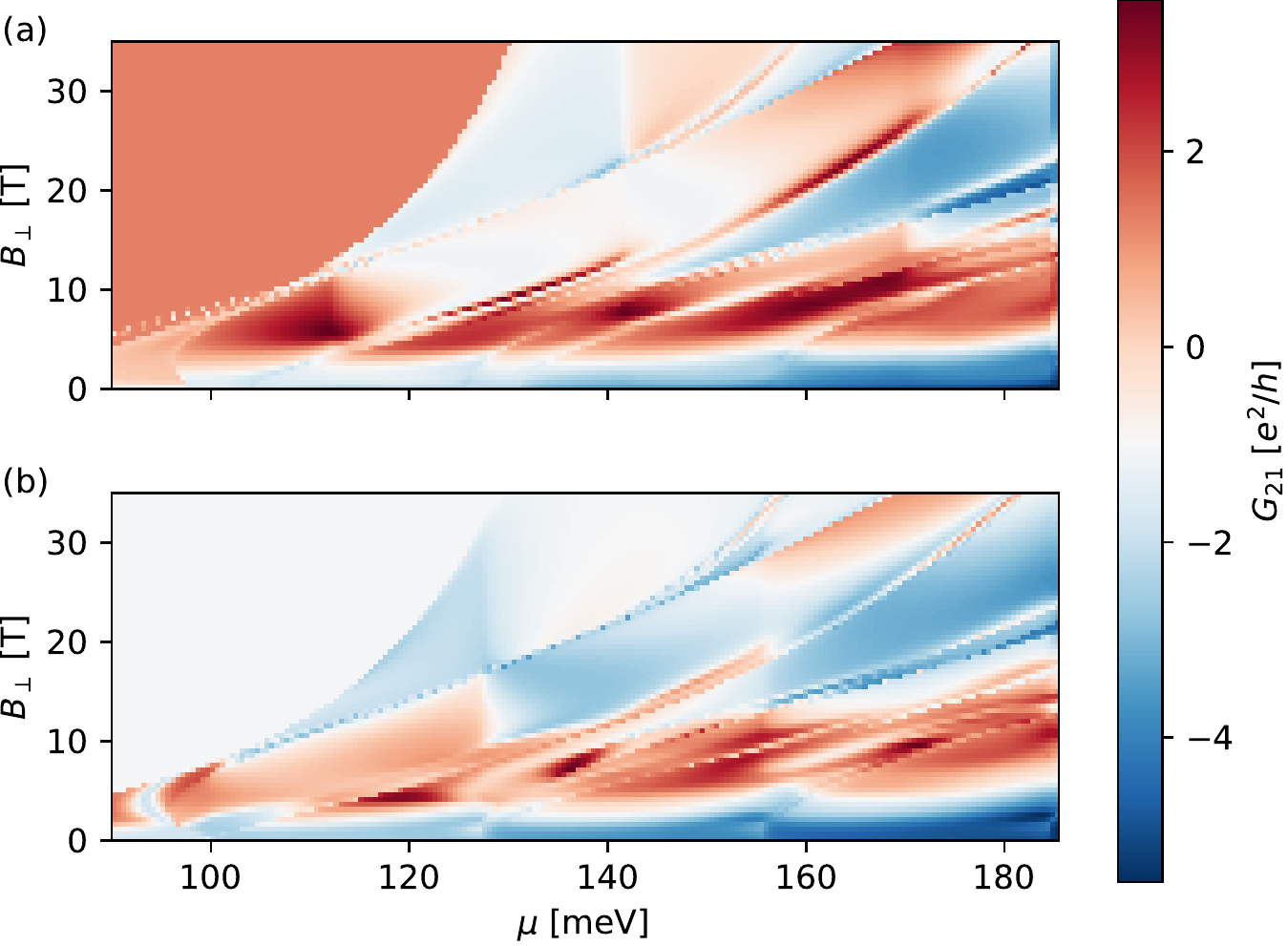}%
\caption{Non-local conductance $G_{21}$ of a T-junction device with a nanowire width of $w=24\,\mathrm{nm}$ and a height of $h=10\,\mathrm{nm}$ as a function of the applied perpendicular magnetic field $B_\perp$ and the chemical potential $\mu$.  In (a) a vortex is introduced at the NS-interface by an axial magnetic field of $B_\parallel=9.58\,\mathrm{T}$ while in (b) no axial field/vortex is present. Red regions correspond to a clear CAR signature $\left(T_{21}^\mathrm{CAR}>T_{21}^\mathrm{T}\right)$, while blue signals stronger normal electron transmission $\left(T_{21}^\mathrm{CAR}<T_{21}^\mathrm{T}\right)$.}
\label{fig:conductivity_map}
\end{figure}

We show this by calculating a 2D density plot of the non-local conductance as a function of the chemical potential $\mu$ and of the perpendicular magnetic field $B_\perp$. 
In Fig.~\ref{fig:conductivity_map} red regions correspond to a positive conductance $G_{21}$, \ie a clear signature that CAR dominates over normal electron transmission. 
For this calculation the T-junction size was reduced to decrease computational costs. 
Below we will revert to the larger system of Sec.~\ref{sec:occurence_of_CAR}, and show that the obtained results also apply for wider nanowires.
The nanowire width and height are respectively $w=24\,\mathrm{nm}$ and $h=10\,\mathrm{nm}$.
The parallel magnetic field was adjusted to give almost exactly a flux $\phi=\phi_0/2$ through the NS-interface with $B_\parallel=9.58\,\mathrm{T}$ in Fig.~\ref{fig:conductivity_map}(a), while in Fig.~\ref{fig:conductivity_map}(b) no axial magnetic field was used. 
The field component parallel to the superconducting lead clearly enhances the CAR signature in a broad parameter range. 
Nevertheless, even without a vortex at the NS-interface, a CAR signature is present in the lower field range over a wide $\mu$ interval. 
For larger systems similar behavior will take place at lower scales of $B_\perp$ and $\mu$.

\begin{figure*}
\includegraphics[width=\linewidth]{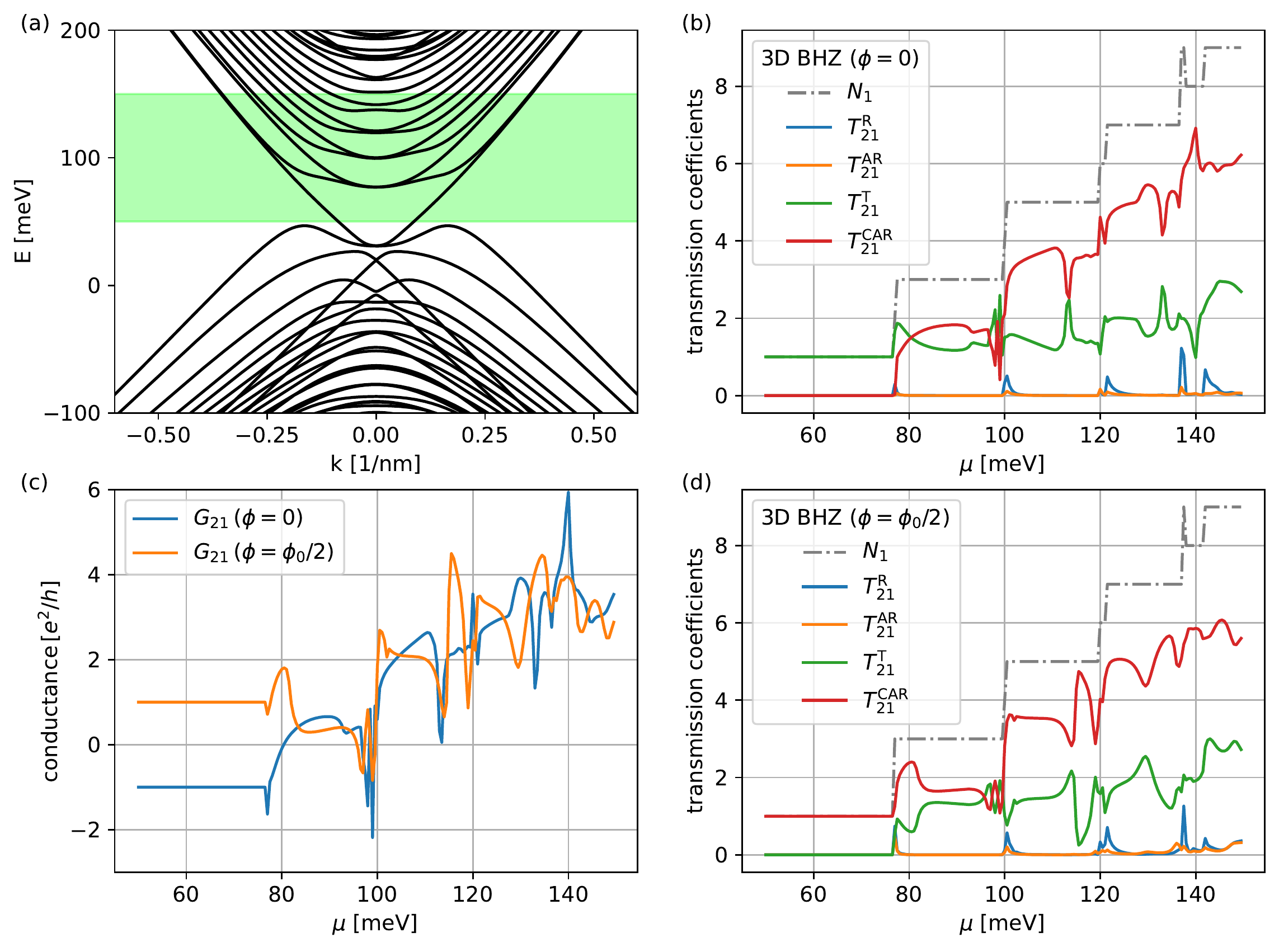}%
\caption{(a) Bandstructure of a TI nanowire with a width of $w=50$ $nm$ and a height of $h=10$ $nm$ in a perpendicular magnetic field of $B_\perp=4$ $T$ calculated for the 3d BHZ model. 
The green shaded region corresponds again to the energy range for our numerical transport calculations.
In (b) and (d) the transmission coefficients for $B_\parallel=0$ and $B_\parallel=4.6\,\mathrm{T}$ are shown.
(c) Non-local conductance calculated with eq.~\ref{eq:G_ba}.}
\label{fig:weak_field_w50}
\end{figure*}

Having established that CAR dominates in a fairly large parameter range, let us switch back to a larger system with a wire width of $w=50\,\mathrm{nm}$.  As opposed to Sec.~\ref{sec:occurence_of_CAR}, where $B_\perp=20\,\mathrm{T}$, we now perform a second calculation at a lower field strength of $B_\perp=4\,\mathrm{T}$. 
Fig.~\ref{fig:weak_field_w50}(a) shows the band structure for this parameter set, 
the green shaded region marking the energy range used in transport calculations. 
Flat Landau levels and the corresponding chiral edge states are starting to form. 
The transmission coefficients for zero axial field ($\phi=0$) and $B_\parallel=4.6\,\mathrm{T}$ ($\phi=\phi_0/2$) are illustrated in Fig.~\ref{fig:weak_field_w50} (b) and (d) respectively.
In the single mode regime switching between no CAR and a robust CAR plateau takes place.  However, and contrary to the strong field case of Fig.~\ref{fig:bise-transmission}, a strong CAR signature survives at higher energies.
This is also clearly observable in Fig.~\ref{fig:weak_field_w50} (c) where the non-local conductance is depicted. 

\subsection{Disordered Systems} 

\begin{figure}
\includegraphics[%
    width=\linewidth,%
]{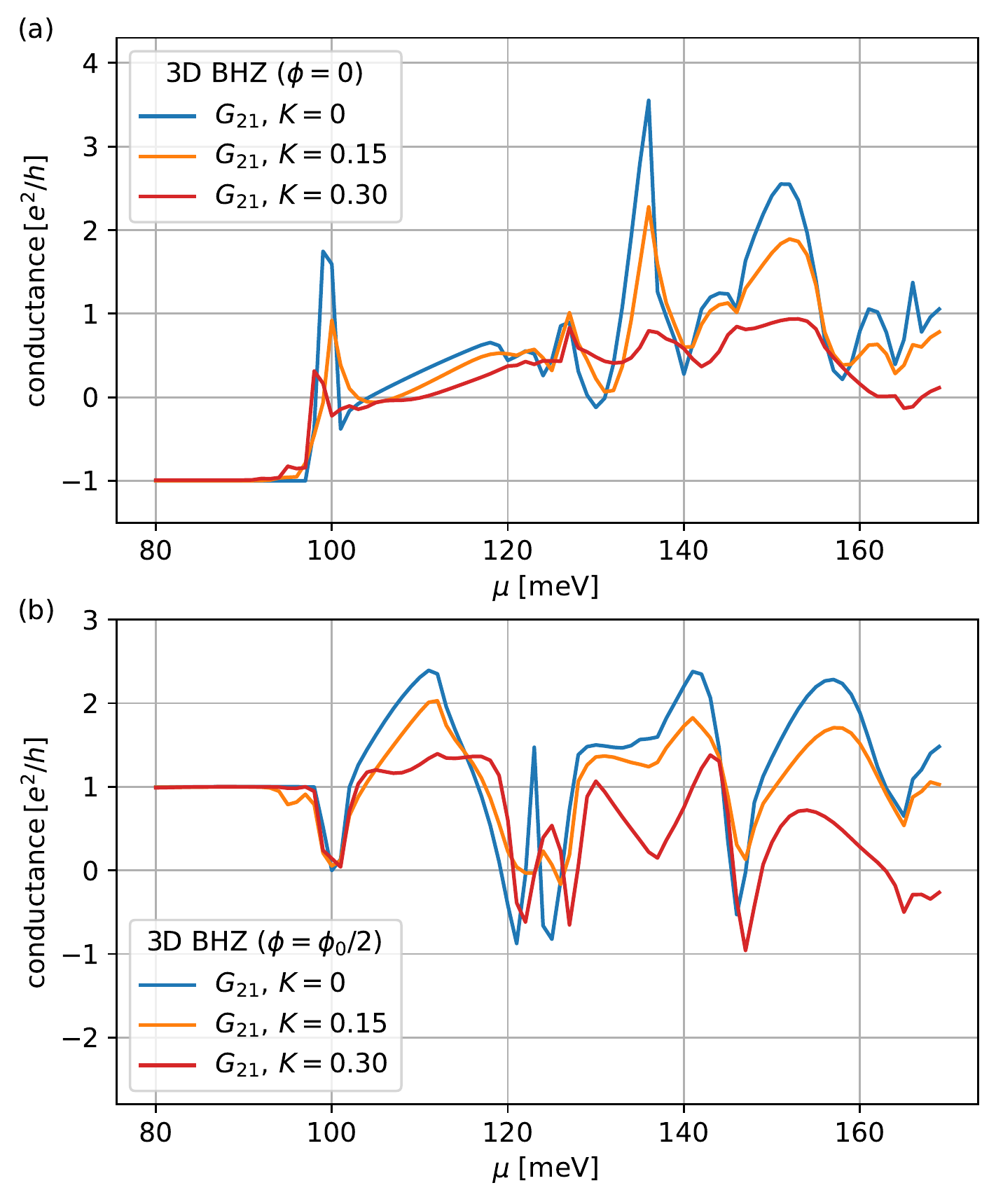}%
\caption{%
    Comparison between the conductance of a clean T-junction and a disordered setup for an axial flux of (a) $\phi = 0$ and (b) = $\phi = \phi_0 / 2$. The disorder conductance was averaged over 20 configurations. The plateau in the single mode regime is robust to the applied disorder.
    \label{fig:disorder_results}%
}
\end{figure}

We test the resilience of CAR signatures to impurities/imperfections, typically present in experimental setups, by performing simulations in disordered samples. 
We use short-range (white noise) disorder\footnote{White-noise disorder is a stronger scattering source than disorder with longer-range correlations.  This is good for the ``stress test'' of CAR, and also appropriate for the relatively small size of our test system.}.
The onsite disorder is chosen from standard normal distribution with amplitude $U = K_0\cdot0.41\,\mathrm{eV}$. Figure~\ref{fig:disorder_results} shows a comparison of the simulation results between the clean and the disordered cases. 
In order to get rid of at least the most significant disorder configuration dependent effects we averaged over 20 disorder sets. 
For the nanowire dimension we chose the same parameters as we used in Fig.~\ref{fig:conductivity_map} and put the perpendicular field to $B_\perp = 7\,\mathrm{T}$.
This allows us to determine if the conductance shown in that density plot is robust to the applied disorder.
The length parameters were $d_n=20\,\mathrm{nm}$ and $d_{sc}=5\,\mathrm{nm}$ so that the NS-interface is lying inside the disordered region. 
Also we put a distance of $45\,\mathrm{nm}$ between the normal leads 1 and 2, so that the incoming modes can possibly scatter already before the NS interface.

By comparing the results of the clean and the disordered simulations one clearly sees that the CAR plateau in the single mode regime is still present. 
At larger energies the disorder is reducing the CAR rate, but also there it is still present. 
The disorder introduces scattering between chiral edge states of the two side surfaces, therefore backscattering and normal electron to electron transmission are enhanced. 
This effect should be reduced in nanowires with a larger width, as this further separates the side surfaces.  
We conclude that in real devices the CAR signature should survive a certain amount of impurities and defects. 

\section{Conclusion\label{sec:conclusion}}


We proposed a device that could be operated as a Cooper pair splitter based on a 3D TI T-junction with one arm in proximity with an $s$-wave superconductor.  The device working principle was studied by examining the inverted process, namely crossed Andreev reflection (CAR), which is tunable by external magnetic fields of moderate magnitude.  Numerical simulations for experimentally relevant parameter ranges (system size, magnetic field strength, disorder) show clear CAR signatures in the transmission coefficients and the non-local conductance.  Signatures can be switched on/off and are more robust in the single-mode regime, which requires stronger fields ($\gtrsim1$ T) and a relatively fine tuning of the electrochemical potential near the Dirac point.  However they are present and fairly disorder-resistant in a wider parameter range.  The Cooper pair splitter in turn should then reliably act as a generator of entangled electron pairs.  \\
On the theory side, we also implemented a 2D effective surface model which is computationally much lighter than a full 3D simulation, and yet produces qualitatively identical transport results.  The 2D model allows treatment of micron-size 3DTI devices, currently computationally too demanding.


\appendix

\section{Matching condition\label{sec:matching}}

On constructing the tight-binding Hamiltonian,
we use the finite differences
\begin{equation}
    \partial_x \psi (x_i)
    \approx \frac{-\imag}{2a} (\psi (x_{i+1}) - \psi (x_{i-1}))
    .
\end{equation}
Thus, the term $\hbar v_F k_x \sigma_y$ in the Hamiltonian
yields the hoppings $t_{i,i+1} = -\imag \hbar v_F \sigma_y / 2a$
from $x_i$ to $x_{i+1}$
and $t_{i+1,i} = \imag \hbar v_F \sigma_y / 2a$, repectively
($a$ beeing the grid spacing $x_{i+1} - x_i$).
At the edge, one has $\psi_1 = U \psi_2$, so that, say,
\begin{equation}
    \psi_2 (x_{i+1}) = U^{\dagger} \psi_1 (x_{i+1})
\end{equation}
and
\begin{equation}
    \partial_x \psi_1 (x_i)
    \approx \frac{-\imag}{2a} (U^{\dagger} \psi_2 (x_{i+1}) - \psi_1 (x_{i-1}))
    .
\end{equation}
Then, the edge hoppings read $t_{i,i+1} = -\imag \hbar v_F \sigma_y U / 2a$
and $t_{i+1,i} = \imag \hbar v_F U^{\dagger} \sigma_y / 2a$, respectively.

As an example,
consider the edge between the $\check{z}$ and $\check{x}$-surfaces.
Then, $\psi_{\check{z}} = U \psi_{\check{x}}$ on the edge where
$U = \exp(-\imag \pi \sigma_y / 4) = (1 - \imag \sigma_y) / \sqrt{2}$
is the spin rotation around the $\check{y}$-axis by $\pi/2$.
Indeed, one finds
\begin{equation}
    U H_{\check{z}}(k_x = k_{-z}, k_y = k_y) U^\dagger = H_{\check{x}}
    .
\end{equation}
On the $\check{z}$-surface,
the finite difference method yields
the hoppings $t_{i,i+1} = -\imag \hbar v_F \sigma_y / 2a$ and
$t_{i+1,i} = \imag \hbar v_F \sigma_y / 2a$ in $\check{x}$-direction.
At the edge $(x,y,z) = (x_e, y_e, z_e)$, one has
\begin{align}
  \psi_i
  &= \psi_{\check{z}} (x_e - a, y_e, z_e)
  \\
  \psi_{i+1}
  &= \psi_{\check{x}} (x_e, y_e, z_e)
  = U^{\dagger} \psi_{\check{z}} (x_e, y_e, z_e)
\end{align}
such that $t_{i, i+1} = -\imag \hbar v_F \sigma_z U / 2 a$ and
$t_{i+1, i} = t_{i, i+1}^{\dagger} = \imag \hbar v_F U^{\dagger} \sigma_z / 2 a$.

\section{Peierl's substitution\label{sec:appendix_peierls}}

For the setup to work efficiently we need magnetic field components. These can be included in the numerical implementation via Peierl's substitution \cite{Peierls1933}
\begin{equation}
t_{x/y/z}=t_{x/y/z}(B=0) \cdot \exp\left(-i\frac{e}{\hbar} \displaystyle\int \scriptstyle \vec{A}\cdot\vec{dl}\right).
\end{equation}
The hopping terms are then modified according to
\begin{align*} 
t_y &= t_y(\vec{B}=0) \cdot \exp\left( \imag \frac{2\pi}{\phi_0}\frac{B_\parallel}{2c} a_z + \imag \frac{2\pi}{\phi_0} B_{\perp} a_x \right), \\
t_z &= t_z(\vec{B}=0) \cdot \exp\left( -\imag \frac{2\pi}{\phi_0}\frac{B_\parallel}{2c} a_y \right), \\
c & \approx 1 - \frac{\langle\lambda\rangle C}{2A_{cs}}
\end{align*}
where C is the circumference of the nanowire and and $A_{cs}$ is the nanowire cross section.
The factor of $c$ is necessary to rescale the flux, as in the 3D model the surface states have a finite extension into the bulk \cite{2018-PhysRevB-Moors_et_al}. The parameter $\langle\lambda\rangle$ is the mean penetration depth of the surface states. In small nanowires the penetration into the bulk will lead to an effective cross section area which is smaller than the actual wire cross section.
For the surface model, $c = 1$.

\begin{acknowledgments}
This work was supported by Deutsche Forschungsgemeinschaft (DFG, German Research Foundation) within Project-ID 314695032-SFB 1277 (project A07) and
the Elitenetzwerk Bayern Doktorandenkolleg ``Topological Insulators''.
We thank R. Kozlovsky for useful conversations.
\end{acknowledgments}

\bibliography{literatur}

\end{document}